\documentclass{svjour3}                     

\usepackage{graphicx,wrapfig}
\usepackage{fix-cm}
\usepackage{mathptmx}      
\usepackage{amsmath,pifont}

\journalname{}

\begin{document}

\title{The Char{\Pifont{psy}C}ive Challenge}
\subtitle{Regulation of global carbon cycles by vegetation fires}
\titlerunning{The Char{\Pifont{psy}C}ive Challenge}
\author{R. Ball}
\institute{R. Ball \at Mathematical Sciences Institute, The Australian National University,
 Canberra 0200 Australia\\
              \email{Rowena.Ball@anu.edu.au, Web: http://www.maths.anu.edu.au/$\sim$ball} }

\date{}

\maketitle

\begin{abstract}
It is an open, but not unanswerable, question as to how much atmospheric CO$_2$  is
sequestered globally by vegetation fires. In this work I conceptualise the question in terms of the general Char{\Pifont{psy}C}ive Challenge, discuss a  mechanism by which
thermoconversion of biomass may regulate the global distribution of carbon between reservoirs,
show how suppression of vegetation fires by human activities may increase the
fraction of carbon in the atmospheric pool, and pose three specific Char{\Pifont{psy}C}ive Challenges of crucial strategic significance to our management of global carbon cycles. 
\end{abstract}

\section{Introduction}
\label{intro}

\begin{wrapfigure}[5]{r}{0.18\textwidth}\vspace*{-6mm}
 \includegraphics[scale=0.4]{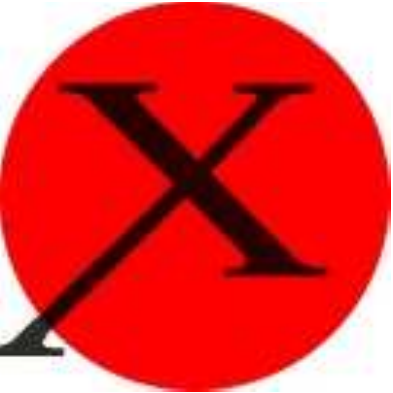}
 \end{wrapfigure}

 The term `char{\Pifont{psy}C}ive' was coined  as a succinct description of the concepts and process of
 sequestering atmospheric carbon in the global
 black carbon reservoir \cite{Ball:2008a}.  It is pronounced ‘tcharckive’---{\Pifont{psy}C} is the Greek letter ‘Chi’. In web and print graphics the visual effect has potentially strong appeal and impact, and as a mnemonic term char{\Pifont{psy}C}ive expresses the long timescales involved: the aim is to  ar\textbf{Chi}ve carbon, i.e., store a valuable substance safely for the long term \cite{charxivearchives}. A somewhat whimsical word-play (cha cha, jive) also suggests the importance of periodic action on short timescales. Indeed char{\Pifont{psy}C}iving is a pretty lively process, as we shall see.

Char{\Pifont{psy}C}iving as a physically meaningful term applies specifically to biochar creation rather than black carbon in general. Biochar, defined as charcoal (syn. char, black carbon or BC) produced by thermal decomposition of biomass, is one of nature's long-term carbon warehouses. It has been made by nature for some 400 million years, for as long as significant terrestrial vegetation has existed~\cite{Taylor:2010}. The outcome of a char{\Pifont{psy}C}iving process is net removal of carbon from the atmospheric pool and storage in the biochar warehouse. Thus, on timescales compatible with human responses to climate change, biomass may be char{\Pifont{psy}C}ived but plastics made from petroleum may not, and used tyres are partially char{\Pifont{psy}C}ive-able. More importantly in the context of the effects of open vegetation fires on global carbon pools, char{\Pifont{psy}C}iving processes include prescribed and traditional burning of forests, grasslands, savannas, and in situ crop residues if managed to enhance charring. Approximate residence times of carbon in burning cycle carbon reservoirs are given in the following table:\\[1mm]
\begin{tabular}{p{7cm}p{5cm}}
 biomass \dotfill &$\sim$1--100 years\\[1mm]
  volatile products of thermal decomposition \dotfill& 0 years \\[1mm]
 atmosphere \dotfill&$\sim$3--200 years\\[1mm]
 biochar \dotfill&thousands of years\\[1mm]             
\end{tabular}

The first Char{\Pifont{psy}C}ive Challenge is this: Can we safely produce, store and distribute enough biochar, over and above that produced by nature, without burning fossil fuels (or otherwise adding to
environmental problems) to significantly lower atmospheric CO$_2$ levels in a time frame compatible with human responses to climate change?  In broad terms, the overall Char{\Pifont{psy}C}ive Challenge deals with scientific, technological, and socioeconomic questions associated with increasing the biochar carbon pool at the expense of the atmospheric carbon pool given that the only way into the biochar carbon pool is through thermoconversion of biomass.

 From this discussion it emerges that global carbon management strategies  involving biochar need to build in  a thorough understanding of the role of fire in regulating carbon cycles.  Vegetation fires, initiated by lightning, humans, or more unusual events such as volcanic eruptions,  are great movers and shapers of the terrestrial environment and significantly influence global carbon balances. Around 8\% of all atmospheric CO$_2$ is chemically reduced by plants yearly. A simple calculation
indicates that, in principle, with current burning rates the mass of carbon currently in the
atmosphere could be decreased by $\sim$4\% over 100 years~\cite{Ball:2008b}. 

This brings us to the second Char{\Pifont{psy}C}ive Challenge, or conundrum at any rate: 
\begin{Pilist}{psy}{183}
\item It is widely accepted that the increasing concentration of  CO$_2$ sequestered in the atmosphere is likely to cause dangerous global warming. 
\item Vegetation fires sequester atmospheric carbon in the global black carbon pool. 
\item The widespread and frequent conflagrations that create biochar in nature are at the very least disagreeable, and often deadly, to humans. Human society---its aspirations, economic and cultural activity---and wildfires cannot coexist in harmony so humans suppress vegetation fires. 
\item Are, then, the imperatives to remove carbon dioxide from the 
atmosphere \textbf{and} suppress wildfires fundamentally incompatible? Have we cornered ourselves? 
\end{Pilist}
This is very much an `in-your-face' question, but in view of current scientific and policy debates on climate change it is one that I believe should be discussed and can be resolved. 

In section \ref{sec2} I describe the mechanism by which the products of biomass thermal decomposition are distributed between char and combustible volatiles, the BioPy thermokinetic oscillator, and discuss its role in maintaining biomass combustion. In section \ref{sec3} I juxatapose the rates of char formation and decay, a facile analysis that leads to the third Char{\Pifont{psy}C}ive Challenge. The three Char{\Pifont{psy}C}ive Challenges presented in this work are summarised in section~\ref{sec4}. 
\section{The BioPy (Bio\textnormal{mass} Py\textnormal{rolysis}) thermokinetic oscillator}
\label{sec2}

\begin{figure}\centerline{
\includegraphics[scale=0.55]{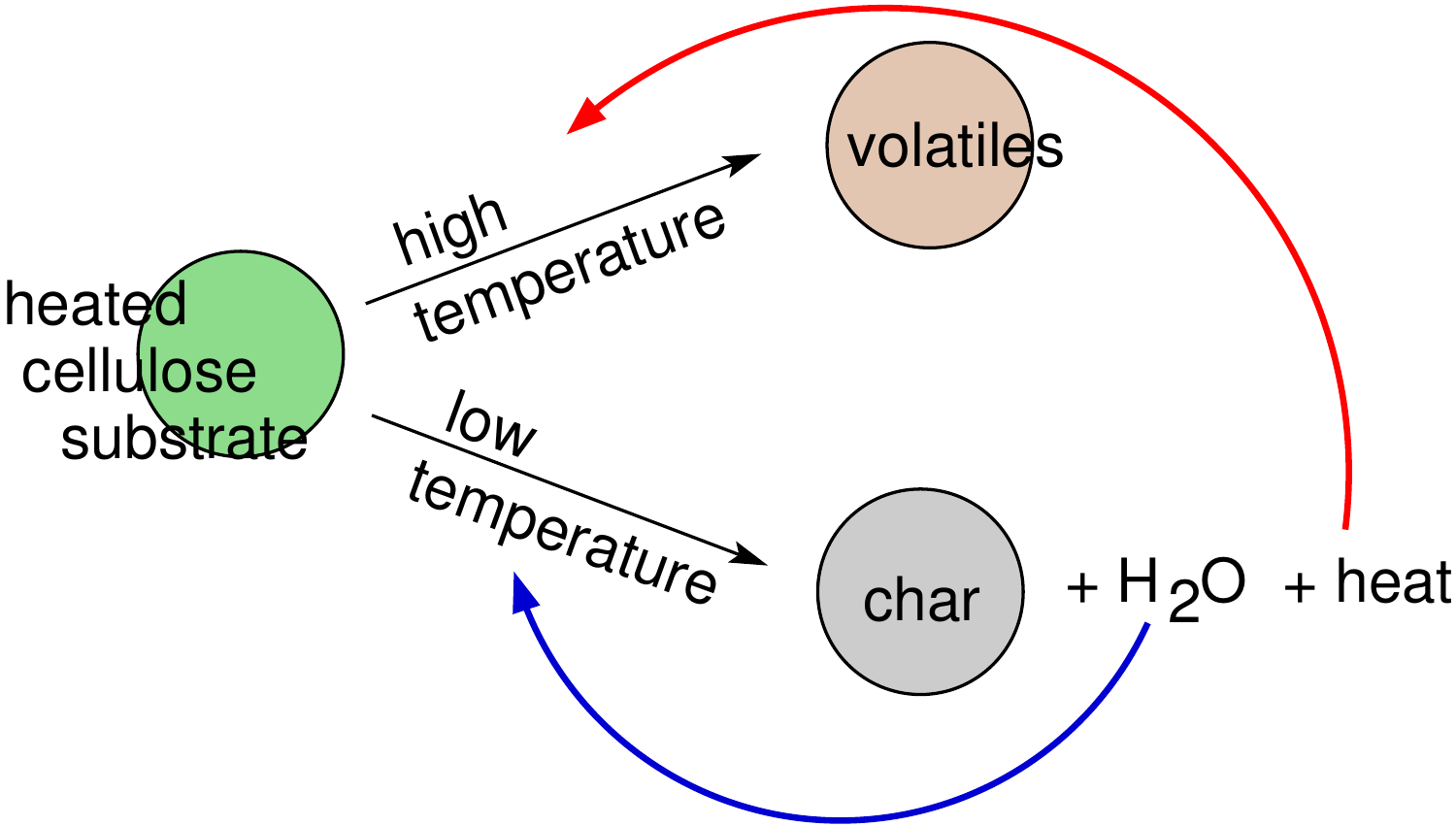}}
\caption{\label{figure1} Chemical feedback autocatalyses the charring path, but the heat produced promotes the competing volatilization process.}
\end{figure}

Biomass combustion is governed by the thermal decomposition chemistry of cellulose, the
major constituent of the terrestrial biomass and by far the most abundant biopolymer on
earth. The detailed organic chemistry of cellulose thermal decomposition will be reviewed elsewhere. Here we consider the thermochemistry (values and signs of the reaction enthalpies) and thermokinetics (temperature (T) dependence of the reaction rate constants, $k(T)=A\exp(-E/RT)$, where $E$ is the activation energy and $R$ is the gas constant) of the two primary processes. In figure~\ref{figure1} these two primary pathways of cellulose thermal degradation are schematized, together with the key thermal and chemical feedbacks. The relevant thermochemical and thermokinetic data for charring and volatilization are given as follows~\cite{Ball:1999}:\\[2mm]
\begin{tabular}{p{0.35\linewidth}p{0.25\linewidth}p{0.3\linewidth}}
\hline
Reaction& Activation energy $E$ & Reaction enthalpy, $\Delta H$\\
\hline
1. cellulose $\longrightarrow $ volatiles & 200--250 kJ/mole& 540 J/g volatiles\\
2. cellulose $\longrightarrow $ char & 110--180 kJ/mole& -2000 J/g char\\
\hline
\end{tabular}\\

Crucial to fire ecology, and to the global carbon cycle as a whole, is the competitive nature
of these two processes: as figure \ref{figure1} indicates, volatiles are produced at the expense of char and vice versa. This reciprocal linkage was originally suggested in \cite{Kilzer:1965} and has been verified by numerous experiments since.

During thermoconversion the instantaneous balance between charring and volatilization is governed by the temperature in the reacting zone. The chemical dehydrations that produce desaturated, highly conjugated precursors to char are low activation energy reactions and therefore set in at relatively low temperatures. The water vapor produced biases the competition towards, or autocatalyses, the charring path. But charring is highly exothermic  overall ($\Delta H$ = -2000 J/g char), because desaturation and crosslinking ensure that more chemical bonds are formed than broken and significant aromatization occurs. The consequently hotter reaction zone allows the  high activation energy volatilization reactions to kick in and take over---for a short time. Volatilization is endothermic, with $\Delta H$ = 540J/g volatiles, and it locally self-damps, thus switching the reaction field again to the charring path. 

It was suggested in \cite{Ball:1999} that this competitive process may provide the mechanism for a thermal oscillator, and in \cite{Ball:2004} it was shown that the solutions of a physicochemical dynamical model for cellulose thermal decomposition obtained using known rate and thermochemical data have the properties of a relaxation oscillator. The model is described in Appendix I and an example time series is given in Figure \ref{figure2}. 
The temperature (a) rises slowly as the charring reactions proceed, as measured by production of water vapor from chemical dehydration of cellulose fragments (b). Heat accumulates in the reaction zone and the temperature spikes because the rate of reactive heat release exponentially overtakes the rate of linear (conductive) heat removal. Just before the temperature jump, the water concentration is at a maximum and volatiles concentration is at a minimum (c). After the jump volatilization begins to cool the system down, the volatiles concentration rises to a maximum causing a temperature collapse, then falls off and the lower temperature charring path takes over again.

\begin{figure}[t]\centerline{
\includegraphics[scale=0.7]{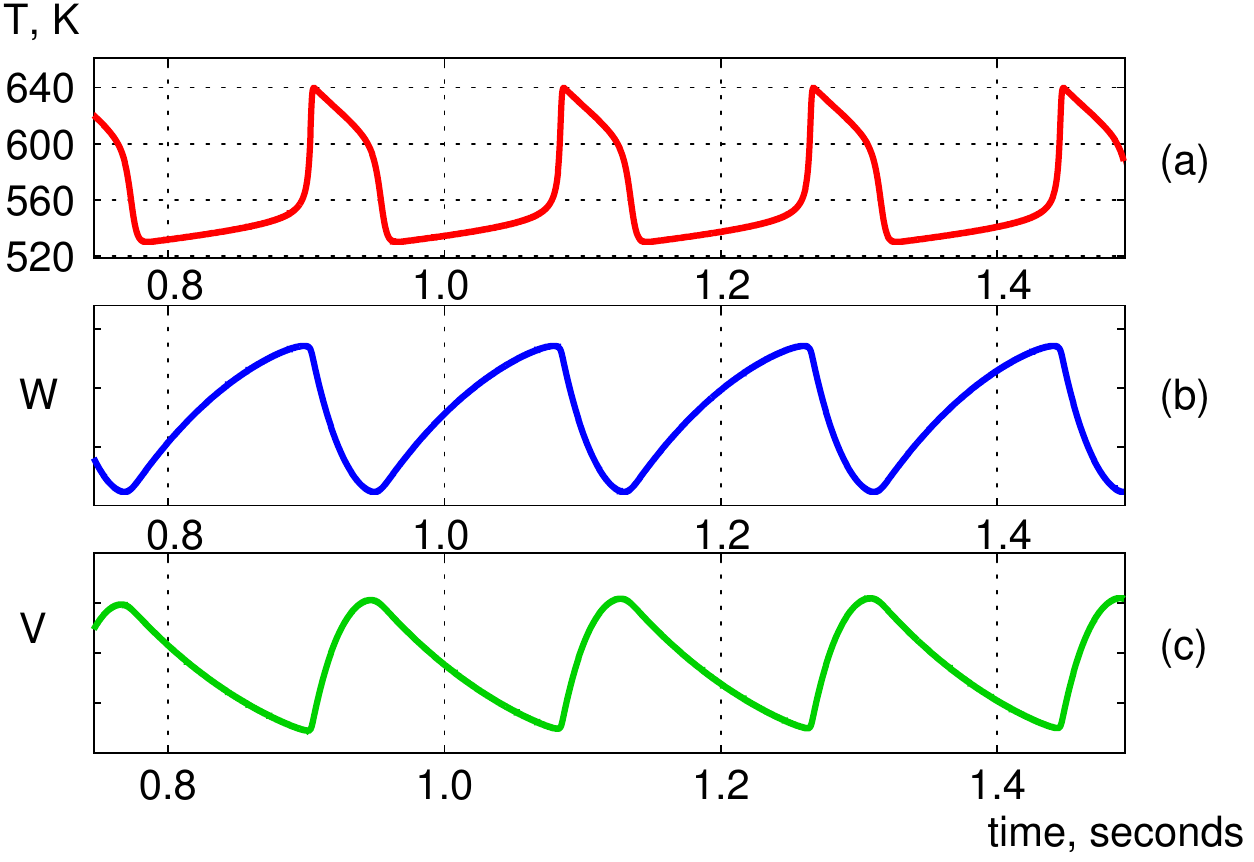}}
\caption{\label{figure2} A time series from the dynamical model for cellulose thermal decomposition in \cite{Ball:2004} shows classic relaxation oscillations. (a) Temperature T in K; (b) W is the water vapour concentration in the reaction zone, a measure of the extent of charring; (c) V is the concentration of volatiles in the reaction zone. W and V are scaled in arbitrary units. }
\end{figure}

This is the BioPy thermokinetic oscillator. Understanding its properties and mechanism are crucial to understanding bushfires \cite{Sullivan:2008} and to engineering biomass pyrolysis. It is related to other well-studied thermokinetic instabilities such as the Sal'nikov and surface reaction thermal oscillators \cite{Gray:1990}, but is distinct in that it features reciprocally coupled endothermic and exothermic reactions.  It is a special case of an Endex thermoreactive system \cite{Gray:1999}, which is described in Appendix II. 

In fact it is not at all difficult to visualize the BioPy oscillator in
action, although few people these days have the apparatus for doing so---a domestic slow
combustion stove. The flames of a wood fire under air control typically flicker, falter, and fluctuate as
the charring and volatilization pathways alternately  dominate, effectively visualizing the pulses of volatiles released by the spikes in temperature. (If you do not have your own wood     stove and spare long winter's evening, try watching in slow motion some of the videos found on YouTube by a search such as `wood fired stove'.) In an open fire the BioPy oscillator tends to be masked by turbulence, and possibly by local fluctuations in oxygen concentration. 

\begin{figure}\centerline{
\includegraphics[scale=0.45]{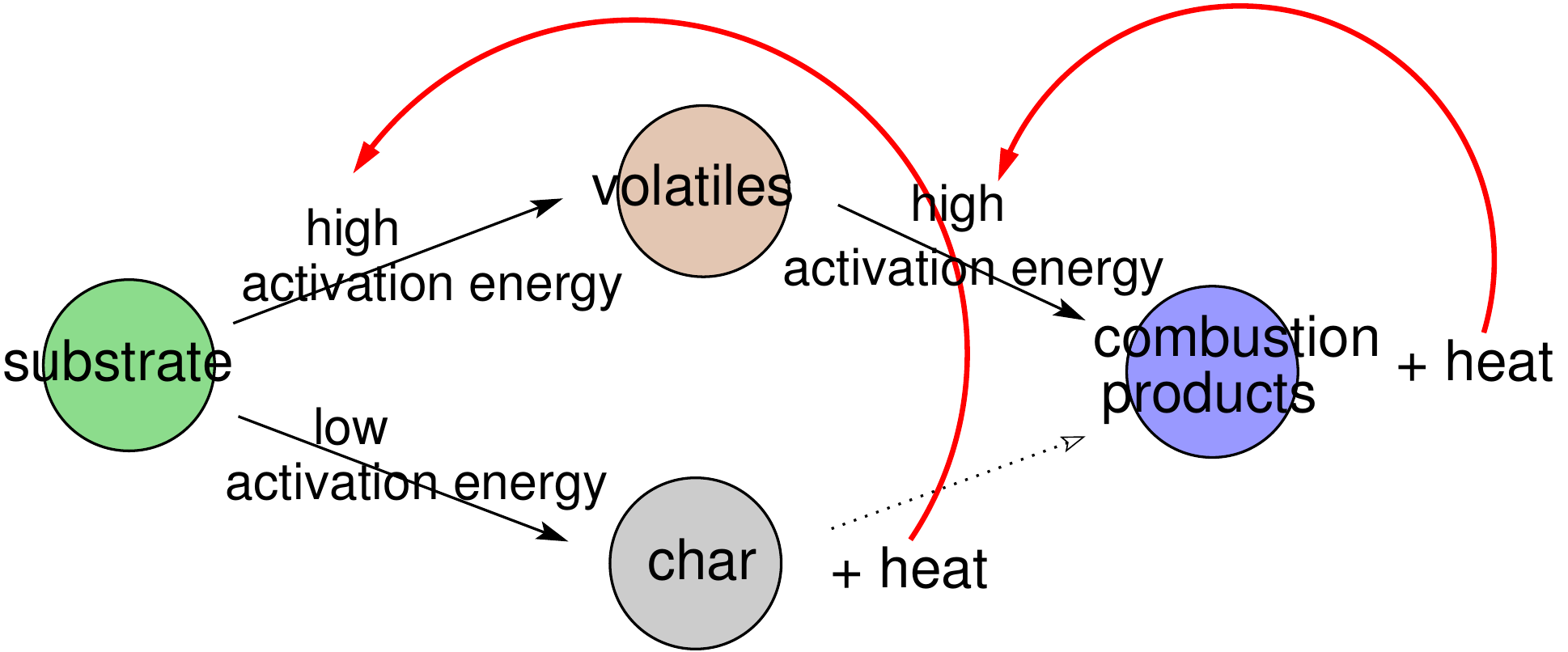}}
\caption{\label{figure3}The heat of charring primarily maintains volatilization of the solid phase and the heat of flaming combustion primarily maintains the gas phase combustion of the volatiles.}
\end{figure}

The flaming combustion path is cartooned in figure \ref{figure3}, where it is indicated that the heat of charring primarily maintains the supply of volatiles from the solid phase substrate and the heat of flaming combustion primarily activates the gas phase combustion of the volatiles. A small fraction of the char is also oxidised during a vegetation fire, as indicated by the dotted arrow. This glowing combustion of the char is a surface reaction that usually extinguishes because of surface diffusion limitations and a high activation energy barrier. 

\begin{figure}\centerline{
\includegraphics[scale=0.45]{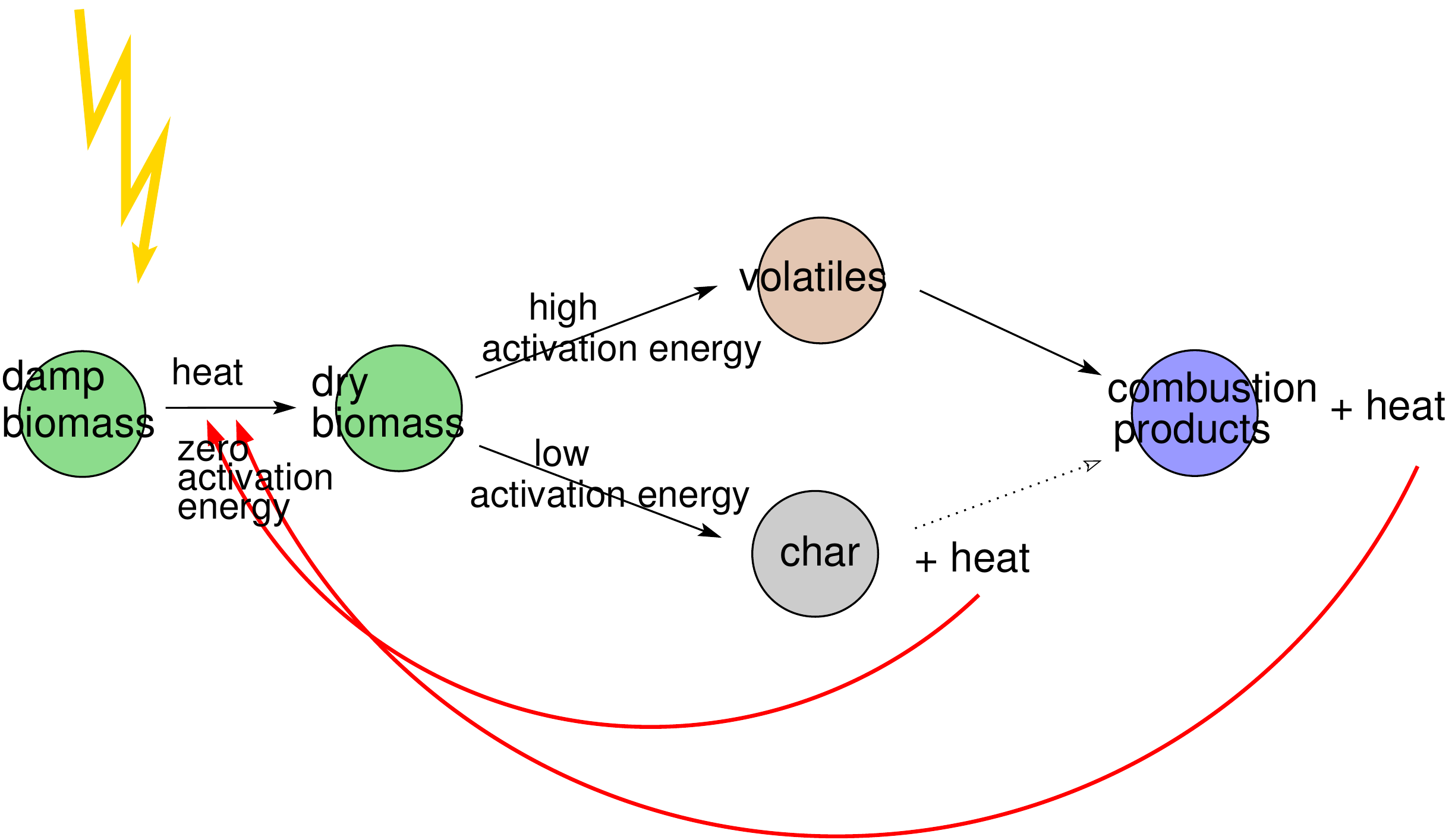}}
\caption{\label{figure4}Combustion of wet vegetation quickly fizzles out because the enthalpy of vaporization of water short-circuits the BioPy oscillator.}
\end{figure}

To appreciate the importance of the BioPy oscillator mechanism in maintaining biomass combustion, consider a (not uncommon) situation where lighting  ignites a mass of very damp biomass, cartooned in figure \ref{figure4}. The heat of charring and heat of combustion preferentially drive the evaporation of water, an endothermic process requiring 41 kJ/mol, because with an activation energy of zero it is kinetically favoured over the high activation energy volatilization and the lower activation energy charring. Flaming combustion is extinguished for lack of volatile fuel to sustain it, the BioPy oscillator ceases to function, and the entire thermoconversion rapidly fizzles out. Viewed in this light, extinction of a vegetation fire by pouring water on it is achieved by short-circuiting the BioPy oscillator.

Thus the BioPy oscillator mechanism as illustrated in figure \ref{figure1} acts to maintain vegetation combustion while fuel is available and the effect is to distribute biomass carbon between the black carbon and  atmospheric reservoirs. In a chronically wet biomass scenario, where combustion cannot be sustained or propagate, refractive char is not formed and nearly 100\% of biomass carbon is returned to the atmosphere as CO$_2$ or methane via the short-term decay path, which uses enzyme-catalysed oxidation reactions of low or zero activation energy. 

\section{\label{sec3}Vegetation fires, biochar and carbon sequestration}

Char accumulates in soils and sediments if its rate of formation exceeds its rate of oxidation. 
How much char is produced globally from wildfire activity per year? What is the rate of char oxidation? 

It is easy to give a rough estimate for the second question: A typical C$^{14}$ activity of charcoal in sediments is around 0.6 that of new wood. The half life of C$^{14}$ is 5730 years, giving the decay time constant as 0.000245/year. For such sedimentary charcoal,  
$\ln (0.6/1.0) = -(\ln(2)/5730) t_{1/2}$, thus $t_{1/2}= 2830$ years,  and literature estimates  are of order 1,300--4000 years \cite{Cheng:2008,Kuzyakov:2009}. 

For the first question, an estimate from a recent analytical review study \cite{Schultz:2008}  for the average annual production of black carbon from wildfires during the 1990s is 2.72 Tg per year.\footnote{The analysis in \cite{Schultz:2008} excludes open burning of agricultural residues. It is unclear whether their estimate refers to aerosol black carbon only or total black carbon. It is useful as a lower bound in any case.} The carbon content of a  typical charcoal is 85\% by mass, giving the mass of carbon sequestered in the black carbon reservoir as 2.31 Tg/year during the 1990s.  

\begin{figure}\centerline{
\includegraphics[scale=0.85]{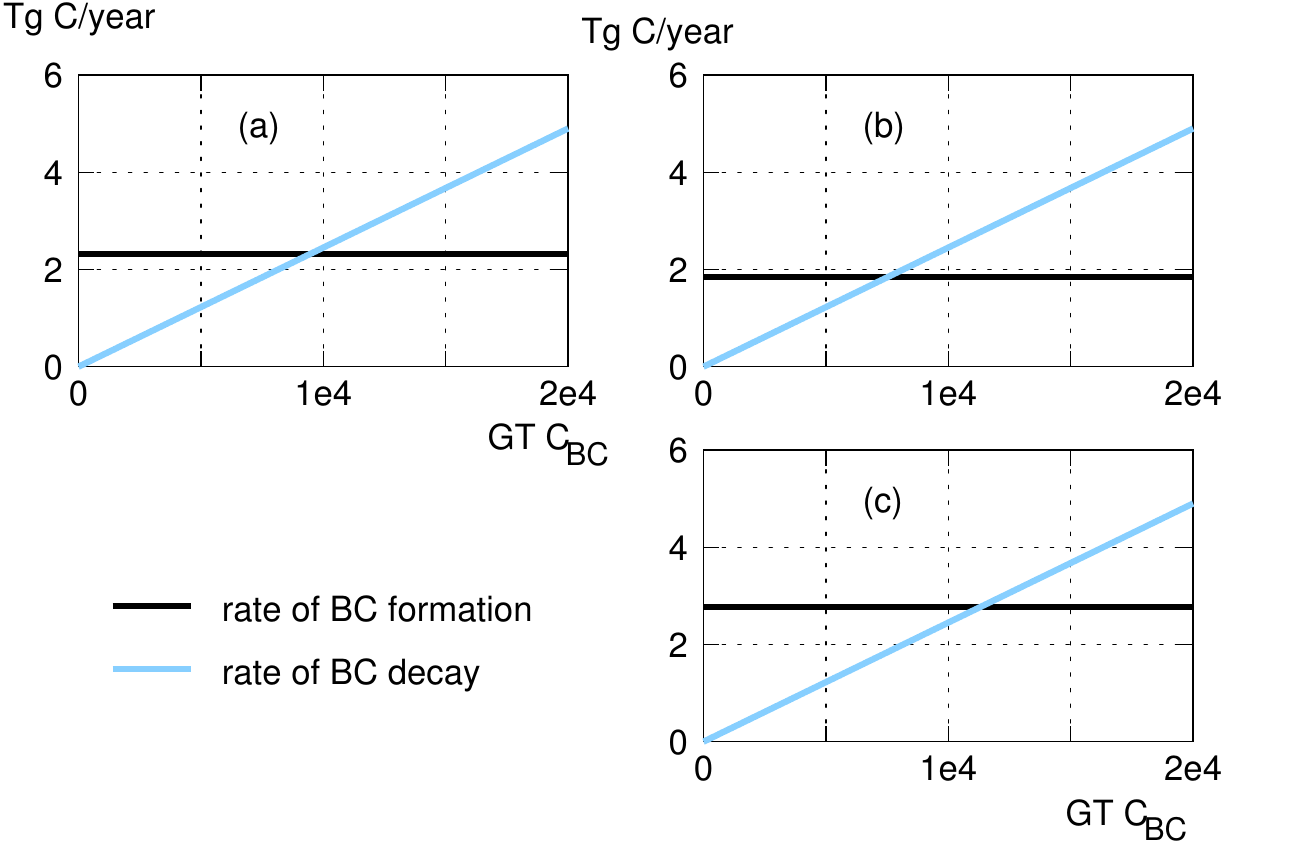}}
\caption{\label{figure5} The rate of BC formation, in Tg C atoms/year, determines the critical global mass of C in the black carbon reservoir, C$_\text{BC}$, above which BC becomes a source of atmospheric CO$_2$.  }
\end{figure}

These rates of BC formation and BC oxidation are plotted in figure \ref{figure5}(a). When the mass of carbon in the global BC reservoir exceeds $\sim$0.95e04\,GT the rate of BC oxidation exceeds the rate of formation and total BC reserves must decrease. In this case BC becomes a source of atmospheric CO$_2$ rather than a sink. 

In figure \ref{figure5}(b) the rate of BC formation is decreased by 20\% from (a). In reality this could be achieved by suppressing open biomass fires, and/or promoting drier hotter fires, and/or by combustion of the char. We see that the critical mass of carbon in the global BC reservoir is significantly lower, $\sim$0.75e04\,GT. 

In figure \ref{figure5}(c) the rate of BC formation is increased by 20\% from (a). This could be achieved by  cooler more frequent open vegetation fire regimes and/or industrial biochar production in globally significant quantities. The critical mass of carbon in the global BC reservoir above which total BC reserves must decrease is considerably greater, $\sim$1.15e04\,GT. 

This brings us to a third Char{\Pifont{psy}C}ive Challenge: How much carbon is currently stored in the global black carbon reservoir? The answer to this crucial question will determine whether human intervention to dramatically escalate biochar production is of any use at all as a CO$_2$ sequestration strategy.

Human activity has promoted fire over recent decades, but suppresses fire in the long term. There are still large gaps and uncertainties in our knowledge of wildland fire emissions, such as yearly mass fluxes, spatiotemporal variability, and longer-term trends. Even the open burning of agricultural residues is poorly quantified. Results of the analytical review study in \cite{Schultz:2008} suggest a significant increase in emissions from wildfires during the period 1960--2000 due to forest and peat soil burning. In the Amazon basin human activity has promoted fire since 1970s \cite{Cochrane:2009}. 
The occurrence of forest fires in Southeast Asia is believed to have increased greatly since the 1960s  and currently prevailing socioeconomic and natural conditions are likely to continue to favour  tropical biomass burning  in the short term \cite{Taylor:2010}.  Yet charcoal records show that global biomass burning declined from 1 to $\sim$1750, and  abruptly after 1870, the latter reduction being due to agricultural and pastoral expansion and fire suppression in intensively farmed areas \cite{Marlon:2008}. 

Human-mediated suppression of biomass burning is likely to continue globally. The widespread fires that have been used over the past few decades  to clear tropical rainforests for the planting of crops and create settled communities with higher standards of living are a transient phenomenon, after which it will be in those communities' interests to suppress biomass burning.

\section{Discussion and conclusions\label{sec4}}

In this work I have argued that wildfires, land management fires, and open burnings of crop residues open the `hatch' that pours carbon into the long-term global reservoir of refractory black carbon. This is a sink for atmospheric CO$_2$ if the rate of BC production (or dimension of the hatch) is greater than the rate of BC decay. The hatch mechanism is the BioPy thermokinetic oscillator, which is effectively short-circuited when wildfires are suppressed. 

The first Char{\Pifont{psy}C}ive Challenge asks the question: Can we produce and distribute enough biochar to stabilise the earth's climate in the short-term? It is the first challenge because the question was posed first, in response to the Virgin Earth Challenge  (www.virgin.com/subsites /virginearth), not because it is the most fundamental problem in the use of biochar to regulate the earth's carbon cycles. 

The second Char{\Pifont{psy}C}ive Challenge asks whether we have painted ourselves into a corner.                                   It asks whether humans'
need to suppress fire is fundamentally incompatible with
nature's use of fire to distribute carbon between long-term
black carbon and short-term atmospheric CO$_2$ pools.

The third Char{\Pifont{psy}C}ive Challenge is, I think, the most important. If we can quantify the amount of black carbon in soils and sediments globally  we can assess whether BC is a source or sink of CO$_2$, then begin to make progress on all the other refinements to our knowledge of carbon flux rates and carbon reservoirs that will be necessary to devise and implement scientifically sound, socioeconomically benign global carbon management.   

\newpage

\appendix
\section*{Appendix I.}

The primary thermal decomposition of a cellulose substrate $S$ occurs via the competing reactions
\begin{align*}
S + \Delta H_1 \quad \overset{k_1(T)}{\longrightarrow} &\quad Z \\
S \quad \overset{k_2(T)}{\longrightarrow} &\quad C + W + \Delta H_2, 
\end{align*}
where $Z$ repesents the primary volatilization product, mostly levoglucosan, $C$ is the char, $W$ is the water of charring, $\Delta H_1$ is the enthalpy of volatilization, $\Delta H_2$ is the enthalpy of charring, and $k_1(T)$ and $k_2(T)$ are the Arrhenius-form rate constants. 

This reaction scheme is incorporated in the following dynamical system:
\begin{align*}
V\frac{dc_W}{dt} &= -VA_1e^{-E_1/RT}c_W + r_W - Fc_W\\
V\frac{dc_Z}{dt}&=VA_2e^{-E_2/RT} - Fc_Z\\
C_{\text{av}}V\frac{dT}{dt}&=V(-\Delta H_1)A_1e^{-E_1/RT}c_W +V(-\Delta H_2)A_2e^{-E_2/RT} +L(T_a-T),
\end{align*}
where $V$ is the volume of the reaction zone, $c_W$ is the concentration of water vapor, $c_Z$ is the concentation of volatiles, $r_W$ is the molar flow rate of  water vapor into the reaction volume, $V/F$ is the residence time of vapor-phase species in the reaction volume, $C_{\text{av}}$ is the averaged volumetric specific heat of the reacting system, $L$ is the rate coefficient of conductive heat transfer from the reaction zone to the environment at temperature $T_a$. It is assumed that volatilization is a (pseudo) zero-order process. A mass balance for $S$ is not included because in this model $S$ is drawn from an infinite pool. This is effectively a two-dimensional dynamical system because $Z$ is determined by $T$, and a linear stability analysis finds that the relaxation oscillations in figure~\ref{figure2} occur where the  eigenvalues of the Jacobian are a complex conjugate pair with positive real parts. 
\section*{Appendix II. What is an Endex system?}

An Endex thermoreactive system consists of an exothermic (heat releasing) reaction that is directly thermally coupled and kinetically matched with an endothermic (heat absorbing) reaction \cite{Ball:1995,Gray:1999,Ball:1999a}. The Endex reactor is the `active' equivalent of a `passive' heat exchanger. In other words the endothermic and exothermic reactions `see' and respond to each other's thermokinetics in real time. The original context for investigation of Endex systems was to achieve intrinsic thermal stability, autothermal operation, full recovery of chemical energy, scaleability, and  co-production of valuable products in chemical reactor systems, and an Endex-configured system has recently been used in a new technology for separating carbon dioxide from flue gas streams \cite{Ball:2010}. 

However, as is often the case with good ideas, it seems that nature can claim prior art with respect to  Endex systems. The thermal decomposition of cellulose appears to be nature's own Endex system, of a special kind in which the competing reactions are intrinsically coupled, that functions to distribute carbon between solid black carbon and atmospheric CO$_2$ reservoirs.

\begin{acknowledgements}
This work is supported by Australian Research Council Future Fellowship FT0991007.
\end{acknowledgements}

\newpage


\end{document}